\documentclass{article}

\usepackage{PRIMEarxiv}

\usepackage[utf8]{inputenc} 
\usepackage[T1]{fontenc}    
\usepackage{hyperref}       
\usepackage{url}            
\usepackage{booktabs}       
\usepackage{amsfonts}       
\usepackage{nicefrac}       
\usepackage{microtype}      
\usepackage{lipsum}
\usepackage{fancyhdr}       
\usepackage{graphicx}       
\usepackage{amsmath}
\usepackage{amsfonts}
\usepackage{array}
\usepackage{mdwmath}
\usepackage{mdwtab}
\usepackage{eqparbox}
\usepackage{enumitem}
\usepackage{url}
\usepackage{cite}
\usepackage{stfloats}
\graphicspath{{media/}}     

\title{Electromagnetic Modeling and Capacity Analysis of Rydberg Atom-Based MIMO System}

\author{Shuai S. A. Yuan $^{1}$, Xinyi Y. I. Xu$^{1}$,  Jinpeng Yuan $^{2}$, Guoda Xie $^{3}$, Chongwen Huang $^{1}$,\\ \textbf{Xiaoming Chen $^{4}$, Zhixiang Huang$^{3}$, and Wei E. I. Sha $^{1,}$}*\\ \\
	$^{1}$ College of Information Science and Electronic Engineering, Zhejiang University, Hangzhou 310027, China.\\
	$^{2}$ State Key Laboratory of Quantum Optics and Quantum Optics Devices, Institute of Laser Spectroscopy, \\Shanxi University, Taiyuan 030006, China.\\
	$^{3}$	Information Materials and Intelligent Sensing Laboratory of Anhui Province, \\Anhui University, Hefei 230039, China.\\
	$^{4}$ School of Information and Communications Engineering, Xi'an Jiaotong University, Xi'an 710049, China.\\
\\ \\
	\texttt{Correspondence: weisha@zju.edu.cn}}

\begin{document}
\maketitle

\begin{abstract} %
Rydberg atom-based antennas exploit the quantum properties of highly excited Rydberg atoms, providing unique advantages over classical antennas, such as high sensitivity, broad frequency range, and compact size. Despite the increasing interests in their applications in antenna and communication engineering, two key properties, involving the lack of polarization multiplexing and isotropic reception without mutual coupling, remain unexplored in the analysis of Rydberg atom-based spatial multiplexing, i.e., multiple-input and multiple-output (MIMO), communications. Generally, the design considerations for any antenna, even for atomic ones, can be extracted to factors such as radiation patterns, efficiency, and polarization, allowing them to be seamlessly integrated into existing system models. In this letter, we extract the antenna properties from relevant quantum characteristics, enabling electromagnetic modeling and capacity analysis of Rydberg MIMO systems in both far-field and near-field scenarios. By employing ray-based method for far-field analysis and dyadic Green's function for near-field calculation, our results indicate that Rydberg atom-based antenna arrays offer specific advantages over classical dipole-type arrays in single-polarization MIMO communications.
\end{abstract}



\section{Introduction}
{W}{ireless} communications have evolved significantly over the years, transitioning from single-antenna systems to multiple-antenna configurations \cite{telatar1999capacity}, then leading to massive \cite{TL2014} and even holographic antenna arrays \cite{Huang2020}. The design of these advanced arrays enables the efficient and skillful utilization of spatial, angular, temporal, frequency, and polarization domains, thereby continuously improving communication performance. However, classical antennas are bounded by many intrinsic limitations, such as Chu's gain-bandwidth limit \cite{chu1948physical} and Hannan's efficiency limit \cite{Hannan1964}, which significantly constrain the potential performance enhancements of current communication systems.

Since it is challenging to overcome these classical limits through conventional physics, Rydberg atom-based receivers \cite{ding2024circularly, yuan2023rydberg,anderson2020atomic,holloway2020multiple,holloway2019detecting,meyer2018digital}, rooted in quantum physics, offer new possibilities by circumventing these restrictions. Rydberg atom-based antennas exhibit several attractive properties, including high sensitivity, broad frequency range, compact size, and low intrinsic noise, making them promising candidates for sensing \cite{anderson2021self,jing2020atomic, liu2023electric,yuan2023quantum,anderson2018vapor,zhang2023quantum} and communication \cite{fancher2021rydberg,anderson2020atomic,yuan2023rydberg,Said2024}. Notably, a single-input multiple-output (SIMO) system has been developed \cite{otto2021data}, and direction-of-arrival (DOA) estimation has been achieved \cite{robinson2021determining}, paving the way for multiple-input multiple-output (MIMO) applications. Additionally, channel models and precoding techniques have also been investigated for Rydberg MIMO communications \cite{gong2024rydberg,cui2024iq,cui2024towards}. From the perspective of antenna engineering, key factors like radiation patterns, efficiency, and polarization must be addressed. However, most existing researches focus on hardware architectures or related signal processing methods, rather than extracting antenna properties from the quantum characteristics of Rydberg atoms. Moreover, the electromagnetic (EM) modeling of such systems remains underdeveloped. A comprehensive approach that combines EM, quantum, and information theories is required to advance the understanding and application of Rydberg MIMO systems.

In this letter, we propose an EM model of Rydberg MIMO systems and perform capacity analysis at both far and near fields, leveraging the two key properties of isotropic reception and lack of polarization multiplexing. Our contributions can be summarized as follows:
\begin{itemize} 
	\item The analysis of a two-level system $S_{1/2} \leftrightarrow P_{1/2}$ is presented to demonstrate why Rydberg systems are insensitive to polarization and direction in MIMO communications, and the distinction between polarization measurement and multiplexing communication is clarified.
	\item The isotropic and polarization effects of Rydberg antenna are considered and integrated into EM-based models for both far-field and near-field communications. Capacity analyses are conducted for revealing the performance of Rydberg MIMO receiver at both far and near fields.
\end{itemize} 

This work unifies the quantum properties of Rydberg antennas with the EM modeling of MIMO communication systems, offering valuable insights for the design of future Rydberg antenna arrays and communication systems.
\section{Principles}
\subsection{Rydberg atom-based antenna array}
\begin{figure}[ht!]
	\centering
	\includegraphics[width=3.5in]{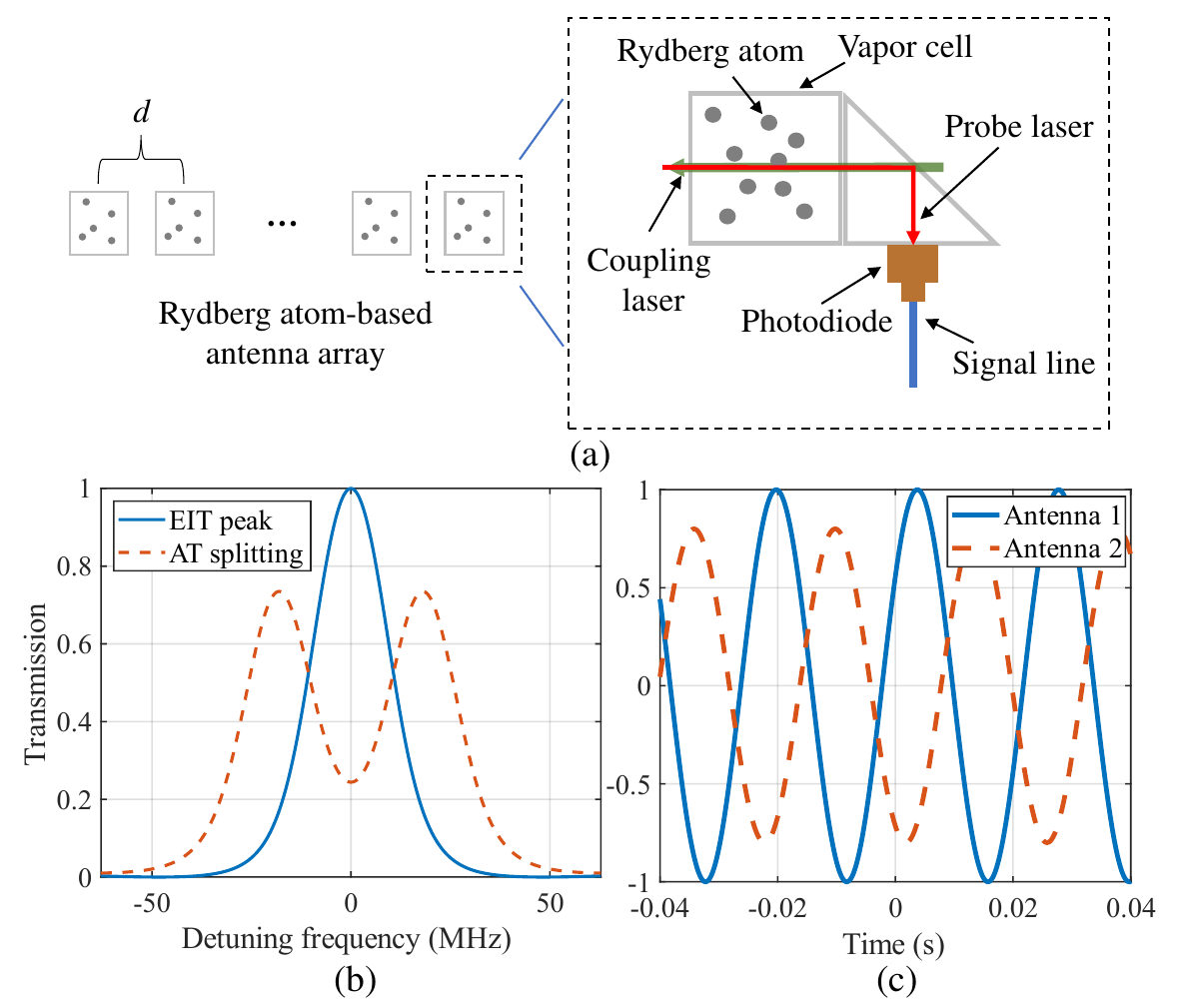}
	\caption{Rydberg atom-based MIMO receivers. (a) Rydberg atom antenna array, the element spacing $d$ is typically smaller than half wavelength. (b) EIT and AT splitting in frequency domain analysis. (c) Time domain signals for retrieving amplitude and phase of the observed field.}
	\label{MIMO system}
\end{figure}
Highly excited Rydberg atoms possess large transition dipole moments to nearby Rydberg states, enabling them to resonate with EM waves in the microwave domain \cite{anderson2020atomic,holloway2020multiple}. Fig. 1(a) illustrates the conceptual design of a Rydberg atom-based antenna array. By using coupling and probe lasers, the Rydberg atoms can detect incoming waves, effectively functioning as atomic antennas. In the frequency domain, electromagnetically induced transparency (EIT) can be established with two laser fields, enabling the Rydberg atoms to become transparent at certain frequencies. The Autler-Townes (AT) splitting effect, induced by microwave fields, facilitates precise field measurements and enables direct encoding and decoding of information \cite{anderson2020atomic}, as shown in Fig. 1(b). Fig. 1(c) demonstrates how amplitude and phase information can be directly retrieved from time-domain measurements, typically with a local oscillator (LO) \cite{robinson2021determining, simons2019rydberg}. Figs. 1(b-c) are computed using the fast algorithm for simulating the Rydberg response developed in our previous work \cite{xu2024fast}. Additionally, because the atomic vapor is typically enclosed in glass, which has minimal scattering of EM waves, mutual coupling between atomic antennas is nearly negligible. For an array element, this results in a near-ideal radiation pattern without distortion and high efficiency.
\subsection{Antenna radiation pattern and polarization}
Intuitively, the Rydberg atoms are conducting stochastic motions in the vapor with no fixed dipole axis, making the receiver insensitive to electric field direction after averaging. To further illustrate this, we consider a system comprising two Rydberg energy levels, $S_{1/2}$ and $P_{1/2}$, each with two sub-levels, $m_j=1/2$ and $-1/2$. The Hamiltonian for the light-atom interaction can be constructed based on this system. According to the transition selection rules, the RF field couple these sub-levels through $\pi$  and $\sigma^{\pm}$  transitions with $\Delta m_j$ = 0 and $\pm 1$. The strength of the individual transitions depends on the orientation of the microwave field's polarization. 
\begin{figure}[ht!]
	\centering
	\includegraphics[width=3.2in]{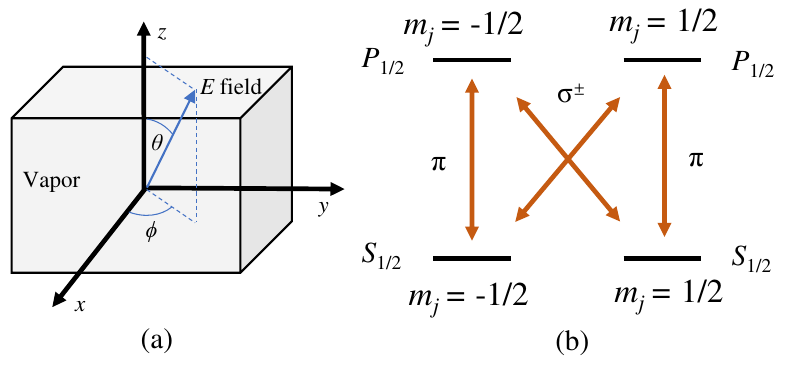}
	\caption{Axis and level diagram for analyzing the effect of incoming wave direction and polarization. (a) Axis and relative angles, $z$ axis is set as the quantum axis, the angle between $+z$ and $E$ field is $\theta$, the angle between $+x$ and the projection of $E$ field on $xoy$ plane is $\phi$. (c) Level diagram with two Rydberg energy levels. The $E$ fields couple each sublevel of $S_{1/2}$ and $P_{1/2}$ through $\sigma^{\pm}$ or $\pi$ transition, dependent on the polarization and direction. }
	\label{MIMO system}
\end{figure}
The axis of this quantum system and the energy levels are depicted in Fig. 2. When the electric field $E$ is along the $z$ axis, the Hamiltonian of this system can be represented by a $4\times 4$ matrix
\begin{equation}
\hat{H}=\frac{1}{2}\left(\begin{array}{cccc}
0 & 0 & -\Omega & 0 \\
0 & 0 & 0 & \Omega \\
-\Omega^* & 0 & 0 & 0 \\
0 & \Omega^* & 0 & 0
\end{array}\right),
\end{equation}
where the $\Omega$ is Rabi frequency, and the entries represent the interactions between the basis states $\left|S_{1 / 2},-\frac{1}{2}\right\rangle$, $\left|S_{1 / 2}, \frac{1}{2}\right\rangle$, $\left|P_{1 / 2},-\frac{1}{2}\right\rangle$, $\left|P_{1 / 2}, \frac{1}{2}\right\rangle$. As depicted in Fig. 2(b), only the $\pi$ transition will appear between $S_{1/2}$ and $P_{1/2}$ with the same $m_j$. With arbitrary rotation of field direction and polarization, the angles between the axis and $E$ can be denoted by $\theta$ and $\phi$ in Fig. 2(a). The contribution of $E$ along $z$ axis accounts for the $\pi$ transition, and the contribution along $xoy$ plane accounts for the $\sigma^{\pm}$ transition with a phase shift $\exp(j\phi)$. Therefore, the Hamiltonian after arbitrary rotation of $E$ field becomes
\begin{equation}
\begin{aligned}
&\hat{H}^{(\theta,\phi)}=\frac{1}{2}\times\\ &\left(\begin{array}{cccc}
0 & 0 & -\Omega \cos \theta & \Omega \sin \theta e^{-j \phi} \\
0 & 0 & \Omega \sin \theta e^{j \phi} & \Omega \cos \theta \\
-\Omega^* \cos \theta & \Omega^* \sin \theta e^{-j \phi} & 0 & 0 \\
\Omega^* \sin \theta e^{j \phi} & \Omega^* \cos \theta & 0 & 0
\end{array}\right).
\end{aligned}
\end{equation}
Solving the eigenvalues of Eqs. (1-2), it can be found that they have the same eigenvalues $\pm \Omega / 2$. The AT frequency splitting $\Delta_\mathrm{AT}$, which can be directly measured in experiments, is equal to the energy difference between two eigenvalues, i.e., $\Delta_\mathrm{AT}=\Omega $. After that, we can obtain the electric field with the relation
\begin{equation}
\Omega=\left|\frac{\mu E}{\hbar}\right|,
\end{equation}
where $\hbar$ is the Planck constant, $\mu$ is the transition dipole moment. It is obvious that the measured splitting effect is not related to the $\theta$ and $\phi$, thus the incoming wave direction and polarization. Similar methods can be extended to more complex systems, and these principles have been experimentally verified \cite{cloutman2024polarization, yuan2024isotropic}. This analysis indicates that the Rydberg atom-based antenna appears an isotropic radiation pattern and is prevented from polarization multiplexing.

\begin{figure}[ht!]
	\centering
	\includegraphics[width=3in]{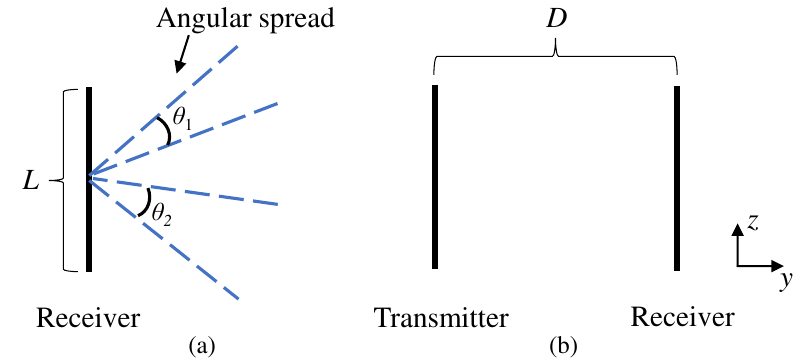}
	\caption{Far-field and near-field communication scenarios, where the sizes of antenna arrays are fixed as $L\times L$, with $N\times N$ antennas uniformly distributed. $L$ is fixed as 5 wavelengths in this paper. (a) The far-field signals can be approximated as incoming plane waves inside certain angular spreads, and an isotropic scattering environment is considered here to match the isotropic pattern of Rydberg antenna. (b) At near field, the transmitting and receiving arrays are directly opposite at a distance $D$, with no scatterers between them. }
	\label{MIMO system}
\end{figure}
It is important to note that shifts or splittings of the field-modified spectroscopic lines depend on the amplitude of the RF field, while the relative strengths of these lines are determined by the orientation of the incident RF field polarization relative to the optical polarization \cite{sedlacek2013atom}. Although this latter property can be employed for polarization measurement, current Rydberg atom systems rely on the shaping of spectroscopic lines for information transmission \cite{anderson2020atomic}. This suggests that polarization multiplexing cannot be used effectively, as the information would not be retrievable in such a configuration.
\section{EM-based modeling of Rydberg MIMO systems}
Based on the previous analysis, Rydberg atom-based antennas function like isotropic point receivers. They can measure the total vector field's amplitude, with equal strength in any direction. For phase, they behave similarly to classical antenna arrays, as demonstrated by their applications in power gain for SIMO systems \cite{otto2021data} and DOA estimation \cite{robinson2021determining}. Therefore, we can construct EM-based models for both far- and near-field scenarios based on these properties.
\subsection{Far-field model}
In the far-field scenario, where considering single polarization is enough, the primary distinction is the isotropic radiation pattern and the absence of mutual coupling. Consequently, far-field performance analysis can be readily conducted using ray-based methods, such as the Kronecker model \cite{xiaoming2013} or the 3rd Generation Partnership Project (3GPP) model \cite{yuan2024breaking}, with appropriate sampling of the radiation patterns. In this letter, we assume an isotropic scattering environment, which aligns with the characteristics of Rydberg antennas due to their isotropic patterns. The correlation coefficient between antennas at positions $m$ and $n$ can be determined with
\begin{equation}
r_{mn}=\frac{\iint  E_{m} (\theta,\phi) E_{ n}^{*}(\theta,\phi)\sin\theta\mathrm{d} \theta \mathrm{d} \phi}{\sqrt{{\smallint\!\!\smallint} |E_{m}(\theta,\phi)|^2 \sin\theta\mathrm{d} \theta \mathrm{d} \phi}  \sqrt{{\smallint\!\!\smallint}  |E_{ n}(\theta,\phi)|^2 \sin\theta\mathrm{d} \theta \mathrm{d} \phi}},
\end{equation}
where $^{*}$ denotes the conjugate operator, $E(\theta,\phi)$ is the far-field radiation pattern. More general forms can be easily included for dual polarizations, cross-polarization discrimination, and non-isotropic scattering environment \cite{xiaoming2013,yuan2024breaking}. After forming the correlation matrix $\mathbf{R}$, the channel matrix considering noise becomes $\mathbf{H}=\left(   \mathbf{H}_w\mathbf{R}^{\frac{1}{2}}\right)$, where the entries of $\mathbf{H}_w$ are i.i.d. complex Gaussian random variables. The transmitting side is considered ideal appearing no correlations with ideal efficiency. Normalization of the channel matrix should be properly made for characterizing the realized gains of the antenna array, especially for dense array \cite{Loyka2009, yuan2024electromagnetic}. Consequently, with power equally allocated, the ergodic capacity is given by
\begin{equation}
C=\mathcal{E}\left\{\log _2\left[\operatorname{det}\left(\mathbf{I}+\frac{\gamma}{N_t} \mathbf{H} \mathbf{H}^{\dagger}\right)\right]\right\},
\end{equation}
where $\mathcal{E}$ denotes the ensemble average, $\mathbf{I}$ is the identity matrix, $\gamma$ is the total signal-to-noise (SNR) ratio, and $N_t$ is the transmitting antenna number. Notice that the Rydberg antenna exhibits significantly lower quantum noise compared to the thermal noise in traditional antennas \cite{fancher2021rydberg}, which generally allows for higher SNR. In this work, however, we assume an equal SNR level for both systems to facilitate a fair comparison of their spatial multiplexing performance
\subsection{Near-field model}
For near-field model, where the vector field should be considered, we can use the dyadic Green's function
\begin{equation}\label{17}
\bar{{{\mathbf{G}}}}\left(\mathbf{r}, \mathbf{r}^{\prime}\right)=\left(\bar{{\mathbf{I}}}+\frac{\nabla \nabla}{k^{2}}\right)g\left(\mathbf{r}, \mathbf{r}^{\prime}\right),
\end{equation}
where $\bar{{\mathbf{I}}}$ is the unit tensor, $k$ denotes the wave number, $\mathbf{r}$ and $\mathbf{r}^{\prime}$ are the positions of receivers and transmitters, $g\left(\mathbf{r}, \mathbf{r}^{\prime}\right)$ is the scalar Green's function. The relation between source $\mathbf{J}(\mathbf{r}^{\prime})$ and field $\mathbf{E}(\mathbf{r})$ can be written in matrix form and
\begin{equation}
\left[\begin{array}{l}
E_x \\
E_y \\
E_z
\end{array}\right]=\left[\begin{array}{lll}
{G}_{x x} & {G}_{x y} & {G}_{x z} \\
{G}_{y x} &{G}_{y y} & {G}_{y z} \\
{G}_{z x} & {G}_{z y} &{G}_{z z}
\end{array}\right]\left[\begin{array}{l}
J_x \\
J_y \\
J_z
\end{array}\right],
\end{equation}
more explicit form of the Green's function can be found in the Appendix II of \cite{yuan2024electromagnetic}. When considering full-polarizations at near field, we can form the channel matrix
\begin{equation}
\mathbf{H}=\left[\begin{array}{lll}
\mathbf{H}_{x x} & \mathbf{H}_{x y} & \mathbf{H}_{x z} \\
\mathbf{H}_{y x} & \mathbf{H}_{y y} & \mathbf{H}_{y z} \\
\mathbf{H}_{z x} & \mathbf{H}_{z y} & \mathbf{H}_{z z}
\end{array}\right].
\end{equation}
For example, $h_{xx}(\mathbf{r},\mathbf{r}^{\prime})=G_{xx}(\mathbf{r},\mathbf{r}^{\prime})$ are the elements of channel matrix $\mathbf{H}_{xx}$ for the transmitters and receivers using $x$ polarization \cite{Shuai2021}. The expressions for other polarizations follow similarly. Unlike classical systems fully described by (8), the Rydberg MIMO receiver does not differentiate between polarizations. Thus, the amplitude of received signal depends solely on the amplitude of total vector field $|E| =\sqrt{|E_x|^2+|E_y|^2+|E_z|^2}$, while the phase term is identical to that of a scalar Green's function, given by $\exp \left(-j k\left|\mathbf{r}-\mathbf{r}^{\prime}\right|\right) $. In a more explicit form, the element for the channel matrix $\mathbf{H}_R$ of Rydberg MIMO is
\begin{equation}
\begin{aligned}
h_R(\mathbf{r},\mathbf{r}^{\prime})&=\exp \left(-j k\left|\mathbf{r}-\mathbf{r}^{\prime}\right|\right) \times\\ &\sqrt{|\sum\limits_{i=x,y,z}G_{xi}|^2+|\sum\limits_{i=x,y,z}G_{yi}|^2+|\sum\limits_{i=x,y,z}G_{zi}|^2},
\end{aligned}
\end{equation}
\begin{figure}[ht!]
	\centering
	\includegraphics[width=3.2in]{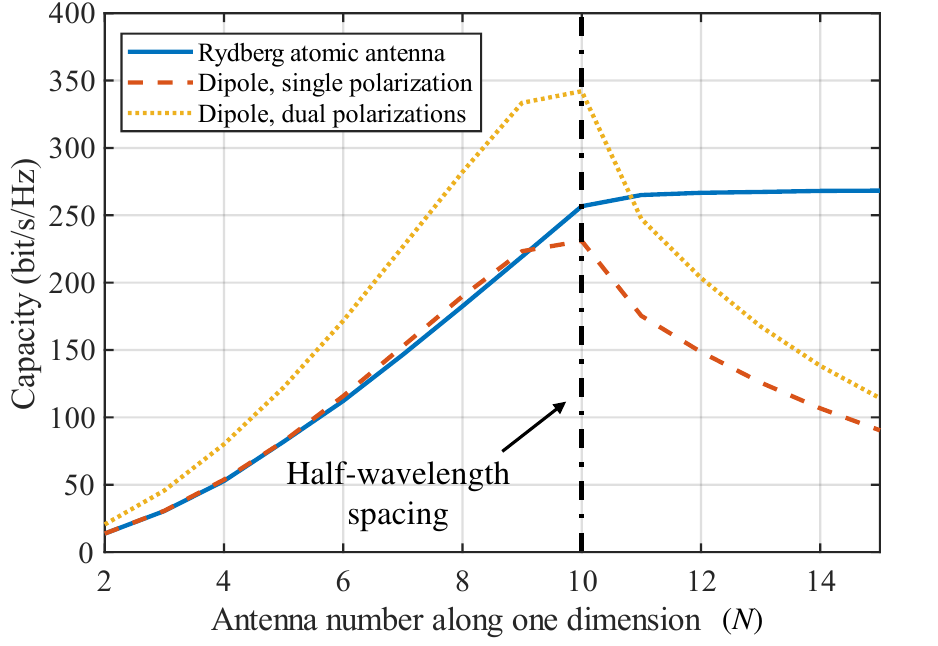}
	\caption{Capacity comparison of MIMO systems build with Rydberg atom and dipole-type antennas at far field (SNR = 10 dB).}
	\label{MIMO system}
\end{figure}\noindent where the $i$ denotes the polarizations used. Typically, a single polarization suffices, as no polarization gain is available. Compared to classical MIMO systems described by $\mathbf{y}=\mathbf{H}\mathbf{x}$, the model for Rydberg ones can be regarded as $\mathbf{y}=|\mathbf{H}\mathbf{x}|\mathbf{P}$, where the $|\cdot|$ here denotes entry-wise absolute value, $\mathbf{y}$ and $\mathbf{x}$ are the receiving and transmitting signals, $\mathbf{P}$ is a diagonal matrix containing the phase term $\exp \left(-j k\left|\mathbf{r}-\mathbf{r}^{\prime}\right|\right) $. For conciseness, the influence of polarization is not explicitly denoted in $\mathbf{H}\mathbf{x}$, but is detailed in (9). For ideal point sources/receivers, $\mathbf{H}\mathbf{x}$ is the same as $|\mathbf{H}\mathbf{x}|\mathbf{P}$ at far field, but not at near field.
\section{Capacity analysis}
\subsection{Far field}
At far field, classical dipole-type antenna appears a cosine-shaped radiation pattern as $E(\theta, \phi) = \cos\theta \exp(\mathbf{k}\cdot\mathbf{r}^\prime)$, where $\mathbf{k}=k\sin \theta\cos\phi\mathbf{x}+k\sin \theta\sin\phi\mathbf{y}+k\cos \theta\mathbf{z}$ is the vector wave number. Hannan's limit \cite{Hannan1964, ShuaiOJAP} can be employed to account for efficiency losses in dense array as $e = \pi S/\lambda^2$, where $S=(L/N)^2$ is the area of a single antenna element, $\lambda$ is free-space wavelength. This efficiency limit concept should also hold for very dense atomic arrays, expressed as $e = 4\pi S/\lambda^2$, since the directivity of each electric-small antenna is 1, unlike the typical value of nearly 4 for traditional antenna. As there is no concept of impedance matching for atomic antennas, this decrease in efficiency can be attributed to insufficient interaction between the atoms and the field, due to smaller atom clusters in vapors. When utilizing dual polarizations, the total power remains fixed, with half allocated for communication in each polarization. In contrast, for Rydberg antennas, an isotropic radiation pattern $E(\theta, \phi) = \exp(\mathbf{k}\cdot\mathbf{r}^\prime)$ is used, and the efficiency losses due to mutual coupling are ignored. 

The capacity results are shown in Fig. 4. The capacity of classical MIMO systems decreases after reaching half-wavelength spacing due to the efficiency loss caused by mutual coupling. While dual polarizations significantly increase capacity, a similar decline at close antenna spacing is still observed. For Rydberg antennas, when considering single polarization, some benefits can still be observed at near half-wavelength spacing due to the higher efficiency and lower correlation between antennas. 
\begin{figure}[ht!]
	\centering
	\includegraphics[width=3.2in]{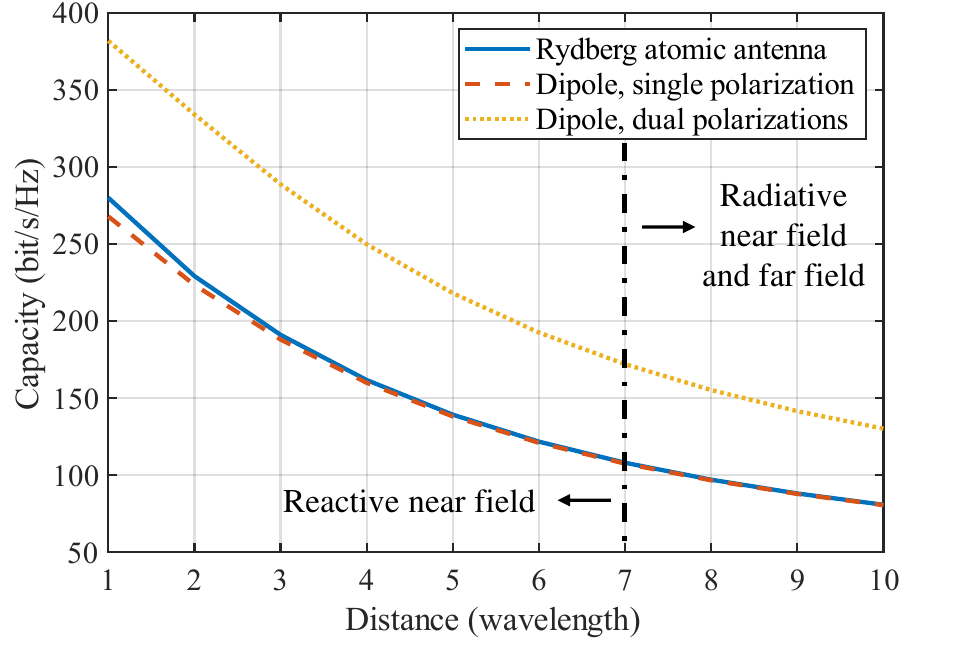}
	\caption{Capacity comparison of MIMO systems build with Rydberg atom and dipole-type antennas at near field, where only single polarization is considered (SNR = 10 dB).}
	\label{MIMO system}
\end{figure}
Moreover, the absence of mutual coupling makes Rydberg antenna a promising candidate for ultra-dense holographic arrays.
\subsection{Near field}
At near-field, we use equation (9) to construct the channel matrix for Rydberg MIMO systems. For classical cases, $\mathbf{H}_{xx}$ from (8) is used for single polarization, while $[\mathbf{H}_{x x},\mathbf{H}_{x y};\mathbf{H}_{y x},\mathbf{H}_{y y}]$ is used for dual polarizations. The element spacing is fixed as half wavelength for both transmitting and receiving sides, while the distances between them is changing. Compared to the channel matrix build with scalar Green's function \cite{Shuai2021}, the main difference in the Rydberg MIMO system lies in the received amplitudes, which are calculated using the dyadic Green's function. The capacity results are shown in Fig. 5, where only minimal benefits are observed at very close distances for the Rydberg MIMO system. As the distance increases, approaching the radiative field region, the performance of both systems becomes nearly identical. Thus, while Rydberg atom-based receivers offer distinctive advantages in sensitivity, size, and noise levels, their spatial multiplexing performance remains similar to that of classical antennas.
\section{Conclusion}
In this letter, we extract the antenna properties from the quantum characteristics of Rydberg atom-based receivers, enabling efficient analysis and seamless integration into existing MIMO system models. Generally, the Rydberg antenna functions as an isotropic scalar point receiver, measuring the amplitude of the total vector electric field, while phase observations behave similarly to those of classical antennas. Compared to conventional MIMO systems, Rydberg-based MIMO communications offer certain advantages in spatial multiplexing performance in some far-field scenarios, while showing similar performance at near field. Given the unique and promising attributes of atomic antennas, such as the absence of mutual coupling, low internal noise, and compact size, further experimental and theoretical studies are warranted to fully explore their potential for MIMO communications.
	\bibliographystyle{unsrt}  
\bibliography{Bibliography}

\vfill


\end{document}